\documentclass[useAMS,usenatbib]{mn2e} 
\usepackage{graphicx}
\usepackage{amsmath} 
\newcommand*\aap{A\&A}

\newcommand*\aj{AJ}

\newcommand*\apj{ApJ}
\newcommand*\apjl{ApJ}

\newcommand*\apjs{ApJS}

\newcommand*\jcap{J. Cosmology Astropart. Phys.}

\newcommand*\mnras{MNRAS}
\newcommand*\na{New A}

\newcommand*\nat{Nature}

\title{The radial acceleration relation in galaxy clusters} 
\author[Chan and del Popolo]{Man Ho Chan$^1$ \thanks{chanmh@eduhk.hk}, Antonino Del Popolo$^{2,3,4}$ \thanks{adelpopolo@oact.inaf.it}
\\ $^1$ Department of Science and Environmental Studies, The Education University of Hong Kong, Tai Po, Hong Kong
\\ $^2$ Dipartimento di Fisica e Astronomia, University of Catania, Viale Andrea Doria 6, 95125 Catania, Italy 
\\ $^3$ INFN sezione di Catania, Via S. Sofia 64, I-95123 Catania, Italy
\\ $^4$ Institute of Astronomy RAS, Pyatnitskaya str., 48, Moscow, Russia}

\begin{document}

\date{Accepted XXXX, Received XXXX}

\pagerange{\pageref{firstpage}--\pageref{lastpage}} \pubyear{XXXX}

\maketitle

\label{firstpage}

\date{\today}

\begin{abstract}
Recently, the discovery of the radial acceleration relation (RAR) in galaxies has been regarded as an indirect support of alternative theories of gravity such as Modified Newtonian Dynamics (MOND) and modified gravity. This relation indicates a tight correlation between dynamical mass and baryonic mass in galaxies with different sizes and morphology. However, if the RAR relation is scale-independent and could be explained by alternative theories of gravity, this relation should be universal and true for galaxy clusters as well. In this article, by using the x-ray data of a sample of galaxy clusters, we investigate if there exists any tight correlation between dynamical mass and baryonic mass in galaxy clusters, assuming hot gas mass distribution almost representing baryonic distribution and that the galaxy clusters are virialized. We show that the resulting RAR of 52 non-cool-core galaxy clusters scatters in a large parameter space, possibly due to our simplifying assumptions and unclear matter content in galaxy clusters. This might indicate that the RAR is unlikely to be universal and scale-independent.
\end{abstract}


\section{Introduction}

The $\Lambda$ cold dark matter ($\Lambda$CDM) model is well known to give very accurate predictions of the observations on cosmological scales, and intermediate scales \citep{2011ApJS..192...18K,Spergel,Kowalski,Percival}. 
To be precise, at cosmological scales the model suffers from the cosmological constant problem \citep{Weinberg, Astashenok}, and the cosmic coincidence problem, and at large scales several tensions are present, between the value of the Hubble parameter, $H_0$, and SNe Ia data, the 2013 Planck parameters \citep{planck} and $\sigma_8$ obtained from cluster number counts and weak lensing. Also the Planck 2015 data are in tension with  $\sigma_8$ growth rate \citep{maca}, and with CFHTLenS weak lensing \citep{raveri} data. Also the small scales ($1-10$ kpcs) are not devoid from issues
\citep[e.g., ][]{moore94,moore1,ostrik,boyl,boyl1,oh,DelPopoloHiotelis2014,DelPopolozz,DelPopolo2017,Zhou:2017lwy,Chang:2018vxs,2019MNRAS.486.1658C}, as the cusp/core problem \citep{moore94,flores}, the ``missing satellite problem" \citep{moore1}, the ``Too-Big-To-Fail problem" \citep{GarrisonKimmel2013,GarrisonKimmel2014}. 
Another issue of the $\Lambda$CDM model is the lack of detection of the dark matter component of the model \citep{Tan,Akerib,Cooley}. Therefore, this long list of problems has ruled out a huge portion of the available parameter space of cold dark matter, especially for the most popular candidate - weakly interacting massive particles (WIMPs). This potentially challenges the standard cosmological model - the $\Lambda$ cold dark matter model \citep{Merritt}. 

%Based on conventional Newtonian dynamics, observations indicate that some missing mass exists in galaxies %and galaxy clusters. It is commonly believed that there exists some unknown particles called dark matter %which can account for the missing mass. As previously reported, the standard dark matter theory can explain %many phenomena from galactic to cosmological scales. However, 
%Recent direct-detection experiments show negative results for dark matter search \citep{Tan,Akerib,Cooley}. Also, 
Recently, the observed radial acceleration relation (RAR) in galaxies show a tight correlation between dark matter and baryons \citep{McGaugh,Lelli,McGaugh2,Lelli2}. The observed scatter is remarkably small and these relations subsume and generalize some other relations, such as the Tully-Fisher relation \citep{Lelli2}. Therefore, the RAR is claimed to be tantamount to a natural law \citep{Lelli2}, which gives an additional challenge to the standard dark matter theory \citep{Merritt}\footnote{Note that \citet{Navia2018}
found that the RAR is broken for galaxies with redshift $z>0.77$}. Here, we want to recall that another scaling law, namely the Mass-Temperature relation in galaxy clusters have been considered as a proof that we need modified gravity \citep{Mota2017}, and later disproved  \citep{DelPopolo}. 

The proposed solutions to the $\Lambda$CDM model issues range from the proposal of modifying the nature of DM \citep{2000ApJ542622C,2000NewA5103G,2000ApJ534L127P,2001ApJ551608S}, or the power spectrum (e.g. \citealt{2003ApJ59849Z}), to delegating the solution to the phenomena related to baryon physics 
(\citealt{1996MNRAS283L72N}; \citealt{Gelato1999}; \citealt{Read2005}; \citealt{Mashchenko2006}; \citealt{Governato2010},\citealt{El-Zant2001,El-Zant2004}; \citealt{2008ApJ685L105R}; \citealt{Cole2011}; \citealt{Saburova2014}).

Another suggestion is that the missing mass in galaxies and galaxy clusters, and the issues of the $\Lambda$CDM model can be explained by modifying the theory of gravity \citep{1970MNRAS1501B,1980PhLB9199S,1983ApJ270365M,1983ApJ270371M,Ferraro2012}, 
%In view of these problems, some suggest that the missing mass in galaxies and galaxy clusters can be %explained by modified gravity 
without the help of dark matter. For example, the earliest version - Modified Newtonian Dynamics (MOND) - suggests that the conventional Newton's second law should be modified when acceleration is smaller than a threshold value \citep{1983ApJ270365M}. Other versions such as Modified Gravity (MOG) {\citep{Moffat}, Galileon Gravity (GG) \citep{Chan}, Emergent Gravity (EG) \citep{Verlinde} and scale invariant theory \citep{Maeder} suggest that some extra terms appear in the gravitational law which can mimic the effect of dark matter. In particular, many theories of modified gravity (e.g. MOG, EG) give a specific connection between dynamical mass and baryonic mass. These theories may be able to account for the observations that the entire shapes of rotation curves of many galaxies trace their baryonic mass distributions.

Nevertheless, recent discovery of ultra-diffused galaxies lacking dark matter (NGC1052-DF2 and NGC1052-DF4) \citep{Dokkum,Dokkum2} and dark matter-dominated galaxy Dragonfly 44 \citep{Dokkum3} may potentially challenge the alternative theories of gravity. Most of the modified theories predict a larger dynamical mass while observations indicate that baryonic mass can nearly fully account for the dynamical mass \citep{Dokkum,Dokkum2}. This result reveals that the canonical stellar mass - halo mass relation \citep{Moster} or the RAR may not be universal. However, some recent studies point out that the discovery of the dark matter-lacking galaxies and the Dragonfly 44 could be consistent with the theory of MOND \citep{Kroupa,Famaey2,Haghi,Haghi2}. Besides, the uncertainties of the distance of these dark-matter-lacking galaxies also make this issue more controversial \citep{Trujillo,Monelli,Haslbauer}. If these galaxies lie at about 10 Mpc, then their stellar masses go down significantly and the ratio between dynamical and stellar mass increases such that these galaxies would become normal dwarf galaxies.   

Furthermore, some recent studies show that the existence of a fundamental acceleration $a_0$ claimed in RAR is rejected at more than $10\sigma$ \citep{Rodrigues}. It means that any alternative to dark matter based on the existence of a fundamental acceleration scale is ruled out. However, recent analyses continue to support the Milgromian acceleration scale \citep{McGaugh4,Kroupa2} and the debate is still on-going \citep{Rodrigues2}.

In this article, we examine the possible RAR for galaxy cluster scale. If there exists a universal fundamental acceleration scale claimed by some modified gravity theories, the RAR should be universal for both galaxies and galaxy clusters. We use the x-ray data of 52 non-cool-core galaxy clusters and show that the RAR is unlikely to be universal. The resulting scatter is too large to be treated as a standard or law-like relation. 

\section{The radial acceleration relation}
Recent fits of the Spitzer Photometry \& Accurate Rotation Curves (SPARC) and some other dwarf spheroidal and early-type galaxies data reveal a strong correlation between the radial dynamical acceleration ($a_{\rm dyn}=v^2/r$) and the radial acceleration predicted by the observed distribution of baryons ($a_{\rm bar}=\partial \Phi_b/\partial r$, where $\Phi_b$ is the gravitational potential of the baryonic component) \citep{McGaugh2,Lelli2}. It is called the radial acceleration relation (RAR), which is a particular form of the mass-discrepancy-acceleration (MDA) relation \citep{McGaugh}. Another similar relation proposed in \citet{Lelli}, the central-surface-densities relation, is also closely related to the RAR. The RAR can be well described by the following empirical function \citep{McGaugh2}
\begin{equation}
a_{\rm dyn}= \frac{a_{\rm bar}}{1-e^{-\sqrt{a_{\rm bar}/a_0}}},
\end{equation}
where $a_0=1.20 \pm 0.02$(random) $\pm 0.24$(systematic) $\times 10^{-10}$ m s$^{-2}$ is a constant. Later, \citet{Milgrom2} shows that the theory of MOND can give a natural explanation to the observed central-surface-densities relation. Also, \citet{McGaugh2} argue that the RAR observed is consistent with MOND's prediction. Generally speaking, the existence of strong correlation between dark matter and baryons favors the alternative theories of gravity \citep{Milgrom2}. 

At the same time, \citet{Desmond,Chan2,Ludlow} show that the standard dark matter theory can also give the RAR. However, the scatter in the RAR cannot be completely reproduced by a cold dark matter (CDM) model \citep{Desmond}. Also, some parameters or feedbacks have to be adjusted to match the observations \citep{Ludlow}. Besides, a recent study using the data of six deep imaging and spectroscopic surveys shows that the intrinsic scatter of the stellar RAR could be as small as 0.11 dex \citep{Stone}. Another uncertainty is that the characteristic acceleration in the RAR could decrease by 40\% with a scatter around 0.12 dex if we include the effect of cold dark baryons \citep{Ghari}. Therefore, this issue is very complicated and it is still controversial to claim the RAR as tantamount to a natural law \citep{Lelli2}. In the following, we explore the RAR in galaxy clusters to see whether it is a universal relation. If it appears with the similar form on a larger scale (galaxy cluster scale), this may give an extra indirect evidence for the claim and favour the alternative theories of gravity. Note that although RAR is consistent with MOND's prediction and it is known that MOND works poorly in galaxy clusters \citep{Ferreras}, it does not mean that galaxy clusters do not have a similar RAR. Here, we are going to examine whether there exists a universal acceleration scale (exhibited in the form of RAR) in both galaxies and galaxy clusters as predicted by some modified gravity theories.

Using the data of galaxy clusters in \citet{Chen}, we can obtain the dynamical acceleration $a_{\rm dyn}$ and the `baryonic acceleration' $a_{\rm bar}$. By assuming hydrostatic equilibrium, the pressure gradient of hot gas in a galaxy cluster is balanced by the gravitational force via the Newton's second law:
\begin{equation}
\frac{dP}{dr}=-\rho_g(r)a_{\rm dyn},
\end{equation}
where $P=\rho_g(r)kT/m_g$, $\rho_g(r)$ is the hot gas mass density and $m_g$ is the average mass of a hot gas particle. Here, the term $a_{\rm dyn}$ is not the real acceleration. It represents the strength of the gravity. Although we are not considering the dynamical acceleration inside a galaxy cluster, the physical meaning of $a_{\rm dyn}$ truly represents the acceleration of a particle if it is placed at a particular $r$ due to gravitational attraction. But in actual situation, the gravitational attraction of hot gas particles are balanced by the gas pressure. If the gas pressure is absent, $a_{\rm dyn}$ is the actual acceleration of the gas particles. In other words, the term $a_{\rm dyn}$ represents the `potential acceleration' due to the gravitational attraction of the dynamical mass in a galaxy cluster. 

For theories of modified gravity such as MOG and EG, Eq.~(2) holds as these theories agree with the Newton's second law. However, the expression of $a_{\rm dyn}$ for modified gravity is different from the conventional Newtonian gravitation $a_{\rm dyn}=GM/r^2$, where $M$ is the total enclosed mass. The only assumption here is using the hydrostatic equilibrium in the calculations. Recent studies point out that using x-ray hydrostatic mass measurements would give only $15-20$\% systematic uncertainty \citep{Biffi}, which is acceptable compared with the measurement errors of the parameters used. 

Observations indicate that the temperature profiles of hot gas in many galaxy clusters are close to constant, except for their inner regions of cool-core clusters \citep{Reiprich,Chen}. Therefore, using a constant temperature profile for each non-cool-core galaxy cluster is a very good assumption. The overall percentage error of mass estimation is less than 15\% \citep{Vikhlinin}. The density profile of hot gas can be described by a $\beta$-model \citep{Cavaliere1976,Cavaliere1978,Chen}:
\begin{equation}
\rho_g=\rho_0 \left(1+ \frac{r^2}{r_c^2} \right)^{-3\beta/2},
\end{equation}
where $\rho_0$, $r_c$ and $\beta$ are fitted parameters. By putting Eq.~(3) into Eq.~(2), we get
\begin{equation}
a_{\rm dyn}=\frac{3\beta kTr}{m_g(r^2+r_c^2)}.
\end{equation}

On the other hand, although the hot gas in galaxy clusters is not really accelerating inward, we define the baryonic acceleration term by
\begin{equation}
a_{\rm bar}=\frac{GM_{\rm bar}}{r^2}=\frac{4\pi G}{r^2}\int_0^r \rho_gr'^2dr'.
\end{equation}
The term $a_{\rm bar}$ as a function of $r$ for each galaxy cluster can be integrated numerically. It represents the gravitational strength due to baryonic mass.

To examine the RAR in galaxy clusters, we choose 52 non-cool-core clusters with $r_c \ge 100$ kpc in \citet{Chen} for analysis. We have ruled out small galaxy clusters because many of them are dominated by the bright cluster galaxies (BCGs) which may induce large systematic uncertainties in hydrostatic mass calculations. By using the parameters obtained in \citet{Chen}, we can get $a_{\rm dyn}$ and $a_{\rm bar}$ as a function of $r$. \footnote{The parameters used are re-scaled because the Hubble parameter assumed in \citet{Chen} is $h=0.5$ while we have used $h=0.68$ in our analysis.} We determine four different positions for each galaxy cluster (at $r=r_c$, $r=2r_c$, $r=3r_c$ and $r=r_{500}$, where $r_{500}$ is the position where average mass density equals 500 times of the cosmological critical density) to calculate the RAR. 

We plot the graph $a_{\rm dyn}$ against $a_{\rm bar}$ to illustrate the RAR in galaxy clusters (Fig.~1). We can see that the resulting RAR for different galaxy clusters scatters in a large parameter space. By comparing with the RAR in galaxies, they have only a very little overlap in the low acceleration regime. Furthermore, the scatter is very large in the small range of $a_{\rm bar}$. The resulting scatter is 0.18 dex, which is larger than the scatter of the RAR in galaxies (0.13 dex) \citep{McGaugh2}. By considering the scatter budget of the involved parameters, the total expected scatter is 0.13 dex (see Table 1). Hence, the intrinsic scatter of the RAR is significant. This suggests that the RAR is not a universal relation and no universal acceleration scale exists. We also fit the data by using Eq.~(1) (see Fig.~2). The best-fit value of the acceleration constant is $a_0=9.5 \times 10^{-10}$ m s$^{-2}$ for the overall empirical RAR, which is much larger than the expected acceleration constant $a_0 \sim 1 \times 10^{-10}$ m s$^{-2}$. This is consistent with MOND prediction in galaxy cluster scale \citep{Sanders,Famaey3}. This may also suggest that there is no universal acceleration scale, or, some additional hot dark matter is needed in galaxy clusters. By plotting the RAR for different subsamples (at $r_c$, $2r_c$ and $3r_c$) in Fig.~2, the corresponding $a_0$ are $1.9 \times 10^{-9}$ m s$^{-2}$ (at $r_c$), $1.2 \times 10^{-9}$ m s$^{-2}$ (at $2r_c$) and $3.9 \times 10^{-10}$ m s$^{-2}$ (at $3r_c$). We can see the trend of larger discrepancies from the galactic RAR as one gets closer to the centres of galaxy clusters. This may indicate the significant effects of astrophysical processes near the centres of galaxy clusters.

\begin{table}
\caption{Scatter budget for the RAR in galaxy clusters.}
 \label{table1}
 \begin{tabular}{@{}lc}
  \hline
  Source &  Residual \\
  \hline
  Errors in $T$ & 0.07 dex\\
  Errors in $r_c$ & 0.06 dex \\
  Errors in $\beta$ & 0.08 dex \\
  Errors in $\rho_0$ & 0.06 dex \\
  \hline
  Total & 0.13 dex \\
  \hline
 \end{tabular}
\end{table}

\begin{figure}
\vskip 10mm
 \includegraphics[width=85mm]{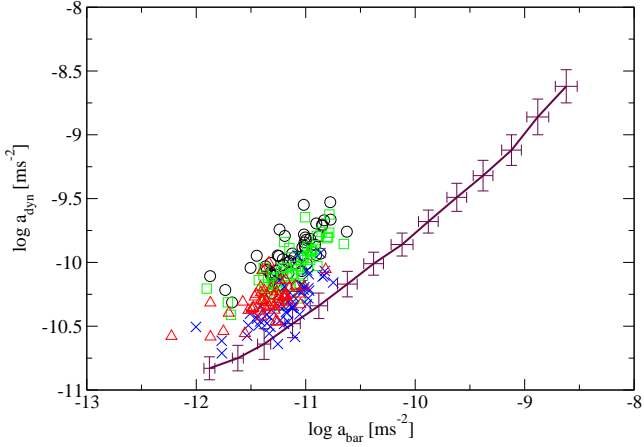}
 \caption{The black circles, green squares, blue crosses and red triangles indicate the RAR for 52 non-cool-core galaxy clusters at $r=r_c$, $r=2r_c$, $r=3r_c$ and $r=r_{500}$ respectively. The data points with error bars are the RAR of spiral galaxies \citep{McGaugh2}.}
\vskip 10mm
\end{figure}

\begin{figure}
\vskip 10mm
 \includegraphics[width=85mm]{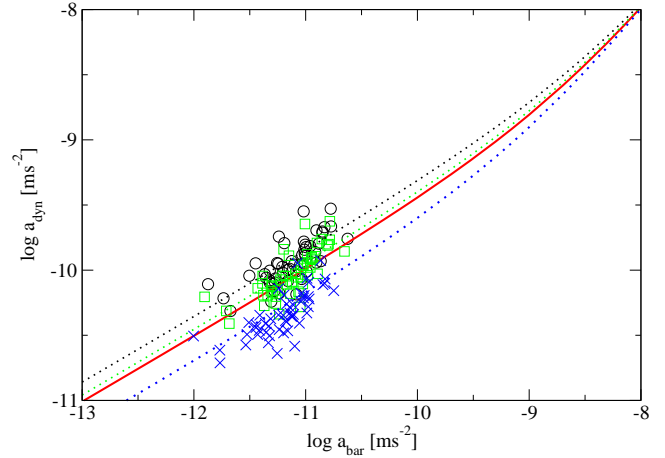}
 \caption{The red solid line indicates the overall empirical RAR following Eq.~(1). The black circles, green squares and blue crosses indicate the RAR for 52 non-cool-core galaxy clusters at $r=r_c$, $r=2r_c$ and $r=3r_c$ respectively. The corresponding coloured dotted lines represent the best-fit RAR at $r_c$, $2r_c$ and $3r_c$ respectively. The best-fit value of $a_0$ for the overall empirical RAR is $9.5 \times 10^{-10}$ m s$^{-2}$.}
\vskip 10mm
\end{figure}

\section{Discussion}

Previous study using 13 galaxy clusters show that the MDA relation or RAR in galaxy clusters is less clear than that for galaxies \citep{Edmonds}. In our study, we investigate a larger sample and show that the resulting RAR for 52 non-cool-core galaxy clusters scatters in a large parameter space. The overall scatter is large and it does not agree with the RAR for spiral galaxies. In other words, the RAR is not a universal relation and it is not scale-independent. Since the RAR gives a specific connection between dynamical mass and baryonic mass, if the RAR is not universal, the connection between dynamical mass and baryonic mass may not be universal either. 

In this study, we assume that the galaxy clusters are virialised and the corresponding hot gas is in hydrostatic equilibrium. Generally speaking, this is a good assumption and the systematic uncertainty involved is about 15-20\% \citep{Biffi}, which is acceptable compared with the scatters. Furthermore, the hot gas mass does not completely represent all baryonic mass in a galaxy cluster. Strictly speaking, some cool gas clouds may exist which could not be detected by x-ray observations. This may make the assumptions about the baryonic distribution less secure. Therefore, if the above simplifying assumptions do not hold, the results in this study may need to revise. On the other hand, the uncertainties of the hot gas temperature profiles and some astrophysical processes like supernovae may also affect the scatters of the RAR in galaxy clusters. More in-depth investigations using simulations or astronomical observations might be required to tackle these issues.

Since there is no universal acceleration scale, the tight RAR in spiral galaxies cannot be treated as an evidence for the alternative theories of gravity. The tight correlation between dynamical mass and baryonic mass in galaxies may be due to some other reasons, such as dark matter-baryon interaction \citep{Famaey} or baryonic feedback processes \citep{Pontzen}. Recent discovery of the dark-matter-lacking galaxy indicates that the correlation between dynamical mass and baryonic mass in galaxies is not necessarily true \citep{Dokkum}. Our result basically supports this new finding, though it is still controversial to claim that the galaxy is completely lacking dark matter \citep{Martin}. 

Although some studies show that RAR might be closely related to MOND \citep{McGaugh2}, our analysis does not have any implication on MOND. It is because MOND does not agree with the conventional Newton's second law when the acceleration is lower than $\sim 10^{-10}$ m s$^{-2}$. Therefore, Eq.~(2) has to be modified in the MOND framework. Nevertheless, it is well known that MOND works poorly in galaxy clusters. Many studies point out that MOND cannot account for the dynamical mass (missing mass) in galaxy clusters \citep{Sanders}. Also, recent radio tracking data of the Cassini spacecraft shows no deviation from General Relativity and it excludes a large part of the relativistic MOND theories \citep{Hees}. Some studies show that the apparent behavior of MOND can be explained by dark matter theory \citep{Dunkel,Chan3}, which means MOND may be just a phenomenological description rather than a universal theory. However, the problem of MOND in galaxy clusters or in cosmological scale might be solved by considering relativistic extensions of MOND (Tensor-Vector-Scalar TeVeS theory). Recent analysis shows that some relativistic extensions of MOND can satisfy the stringent constraints of the speed of gravitational waves \citep{Skordis}. Some other studies point out that the problem of MOND in galaxy clusters could be alleviated if there exists some massive sterile neutrinos in our universe \citep{Angus}. Adding sterile neutrinos to MOND might be able to account for the RAR in galaxy clusters because the `baryonic mass' could be increased by a certain factor such that the empirical RAR would be fitted to the galaxy cluster data with the canonical value of $a_0$. The possibility of the existence of massive sterile neutrinos is hinted by some ground-based experiments \citep{Aguilar} and now becoming a hot issue in particle physics. It is also possible to have some connections between MOND and long-range quantum gravity \citep{Cadoni}. Other versions of MOND like ``extended MOND" (EMOND) in which the acceleration scale depends on the depth of gravitational potential can explain the dynamics of galaxy clusters \citep{Zhao}. Therefore, MOND is still a popular alternative theory of gravity currently. Further astronomical observations or cosmological studies are required to settle this controversial issue.

\section{acknowledgements}
We are grateful to the referee for helpful comments on the manuscript. This work was supported by a grant from the Research Grants Council of the Hong Kong Special Administrative Region, China (Project No. EdUHK 28300518) and the grants from the Education University of Hong Kong (RG2/2019-2020R and RG7/2019-2020R).

%\bibliographystyle{mn2e}
%\bibliography{Burkert}

%\end{document}

\label{lastpage}

\end{document}